\documentclass[twocolumn,superscriptaddress,amsmath,amssymb]{revtex4-2}
\usepackage{color}
\usepackage{graphicx}
\usepackage{subfigure}
\usepackage{booktabs}
\usepackage{dcolumn}
\usepackage{bm}
\usepackage{amsmath}
\usepackage{verbatim}
\usepackage{lineno}
\usepackage[T1]{fontenc}
\usepackage{mathtools, leftindex}
\usepackage{ulem}
\usepackage{subfiles}

\newcommand{\beq}{\begin{eqnarray}}
\newcommand{\eeq}{\end{eqnarray}}

\begin{document}




\title{Supercritical-subcritical correspondence, asymmetric effects and antisymmetric corrections near a critical point}
 

\author{Xinyang Li}
\affiliation{Peng Huanwu Collaborative Center for Research and Education, Beihang University, Beijing 100191, China}

\author{Yuliang Jin}
\email{yuliangjin@mail.itp.ac.cn}
\affiliation{Institute of Theoretical Physics, Chinese Academy of Sciences, Beijing 100190, China}
\affiliation{School of Physical Sciences, University of Chinese Academy of Sciences, Beijing 100049, China}
\affiliation{Wenzhou Institute, University of Chinese Academy of Sciences, Wenzhou, Zhejiang 325000, China}

\date{\today}

\begin{abstract}
The second-order phase transitions in the Ising model and liquid-gas systems share a universality class and critical exponents, despite the absence of $Z_2$ symmetry in the liquid-gas Hamiltonian. This discrepancy highlights a central puzzle in critical phenomena: what is the influence of asymmetry on scaling laws? For over a century, this question has been explored through examining violations of the empirical ``rectilinear diameter law'' for the subcritical coexistence curve, where asymmetry could generate singular corrections. Here, we extend this investigation to the supercritical regime. We propose a supercritical-subcritical correspondence, drawing a formal analogy between the subcritical coexistence curve and recently defined supercritical boundary lines ($L^\pm$ lines). Our theory predicts that the linear mixing of physical fields - a hallmark of asymmetric systems - produces universal scaling corrections, with antisymmetric coefficients, in these supercritical loci.
We verify these predictions using  liquid-gas data from the NIST database and a model liquid-liquid transition. Furthermore, we demonstrate that the same asymmetric scaling framework governs the behavior of higher-order cumulants in the order parameter distribution.
\end{abstract}

\maketitle

{\bf Introduction.}
The notion of universality is well established for critical phenomena – systems belonging to the same universality class share the same critical exponents, independent of microscopic details.
However, such systems do not need to present the same  symmetry in their Hamiltonians. A paradigmatic example is the comparison between the ferromagnetic-paramagnetic phase transition in the Ising model and the liquid-gas critical point, both of which belong to the Ising universality class~\cite{Fisher1983}. The Ising Hamiltonian has an obvious $Z_2$ symmetry because the energy is invariant if all spins are flipped. In the case of the liquid-gas critical point, the elementary lattice gas model enjoys the same $Z_2$ symmetry -- termed a ``particle-hole symmetry'' -- since it can be directly mapped to the Ising model~\cite{lee1952statistical}. However, such a symmetry  disappears if the particles are allowed to move continuously in the space, or the Hamiltonian is redefined to be asymmetric in the lattice gas model~\cite{mermin1971lattice}. 
In general, realistic liquid-gas systems conform the Ising universal critical exponents without  generic  $Z_2$ symmetry  -- for this reason, such systems are often called ``asymmetric''. In fact, asymmetric  systems are ubiquitous in the nature; most phase transitions (liquid-gas, liquid-liquid, etc.) in atomic and molecular systems are asymmetric. 

\begin{figure}[!htbp]
  \centering
  \includegraphics[width=0.85\linewidth]{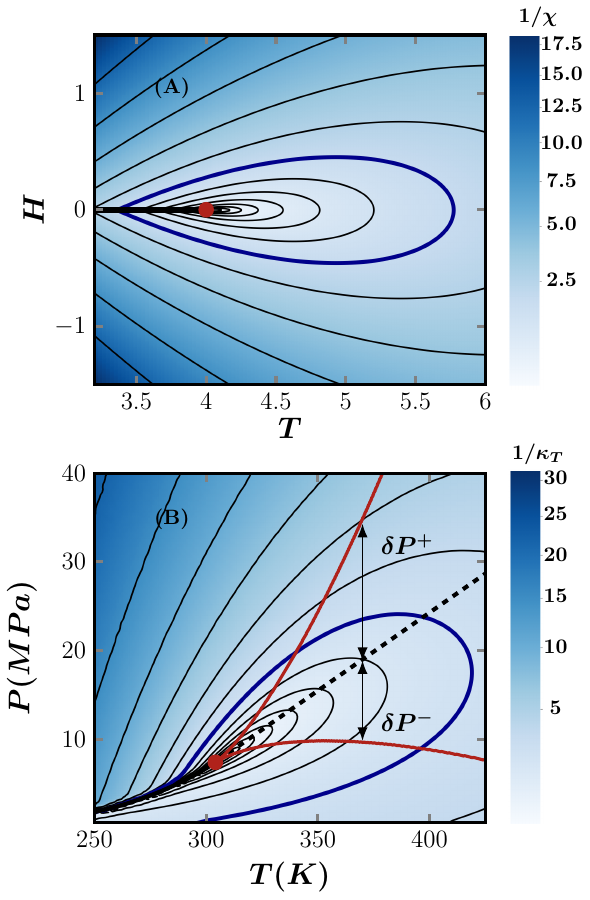}
  \caption{{\bf Phase diagrams and EOSs for symmetric and asymmetric systems.} (A) $H-T$ phase diagram of the 2D mean-filed Ising model. 
  (B) $P-T$ phase diagram of $\rm{CO_2}$, obtained from the NIST database~\cite{NIST}.
  The thick black line is the first-order transition (coexistence) line, and the red circle the critical point. 
 The thin contour lines are constant order parameter EOSs, with one typical EOS highlighted by a thicker line.
  In (B), the dashed black line is the  critical isochore ($\rho=\rho_{\rm c}$), and   the red lines are $L^{\pm}$ lines~\cite{li2023thermodynamic}. 
  }
  \label{fig:Z2_illustration}
\end{figure}

In asymmetric systems, the Hamiltonian  $Z_2$ symmetry
is naturally  reflected in the equation of state (EOS). 
Without loss of generality, we consider the constant susceptibility lines  in the external field-temperature phase diagram. 
In the magnetic field - temperature ($H-T$) phase diagram of the Ising model (Fig.~\ref{fig:Z2_illustration}), a typical contour line, along which the magnetic susceptibility $\chi = (\frac{\partial m}{\partial H})_T$  is a constant, is obviously symmetric with respect to the first-order transition coexistence line  ($H=0$). Contrarily, in the pressure - temperature ($P-T$) phase diagram of the liquid-gas system, a contour line with a constant  compressibility {$\kappa_T =  (\frac{\partial \rho}{\partial P})_T$}   does not have an apparent  symmetry. In particular, such a line does not close a loop below the critical point. A key unresolved problem is the explicit role of asymmetry in the behavior of EOS near the critical point.

The asymmetric effects in the EOS have been indeed intensively discussed, with particular attention paid to the violation of the famous {\it law of rectilinear diameter}~\cite{cailletet1886recherches}, 
 which asserts a linear dependence  of the mean liquid-gas density (``diameter'' of the coexistence curve) $\rho_{\rm d} = \frac{1}{2}(\rho_{\rm L} + \rho_{\rm G})$ on $\Delta \hat{T} \equiv (T- T_{\rm c})/T_{\rm c}$ near the critical temperature $T_{\rm c}$, 
\beq
\rho_{\rm d} =  \rho_{\rm c} + C |\Delta \hat{T}|,
\label{eq:diameter}
\eeq
where $\rho_{\rm L}$ and $\rho_{\rm G}$ are the liquid and gas densities at the same $T$,
$\rho_{\rm c}$ the critical density, and $C$ a system-dependent coefficient.
Historically, this law played a very important role in determining the critical density~\cite{reif2010history}.
The Ising model has $C=0$, which is a special case of a linear relationship. For asymmetric systems, many theories have proposed a ``singular diameter''~\cite{rehr1973revised, rehr1973revised, mermin1971lattice,  wilding1992density,  mermin1971generality, widom1970new, hemmer1970fluids, ley1977revised, nicoll1981critical, anisimov2006nature,PhysRevE.75.051107}, 

\beq
\rho_{\rm d} =  \rho_{\rm c} + D_1 |\Delta \hat{T}|^{2\beta}
+ D_2 |\Delta \hat{T}|^{1-\alpha}
+ \cdots,
\label{eq:diameter2}
\eeq
where $\alpha \approx 0.11$ and $\beta \approx 0.33$ are standard critical exponents (3D Ising).
The expression shows  $d\rho_{\rm d}/dT$ diverging weakly near the critical point. Naturally, one would expect such singular correction terms to be absent if the coexistence curve is symmetric with respect to the liquid and gas sides.
Experimental examinations on the  asymptotic critical behavior of $\rho_{\rm d}$ are challenging~\cite{pestak1987three, hahn2004high, anisimov2006nature, singh1990rectilinear, huang2005density, nowak1997measurement}. 
An important reason is that the subcritical ($T<T_{\rm c}$) scaling regime for Eq.~(\ref{eq:diameter2}), whose size could be system-dependent,  is too narrow to be detected in many systems.

\begin{figure}[!htbp]
  \centering
  \includegraphics[width=0.75\linewidth]{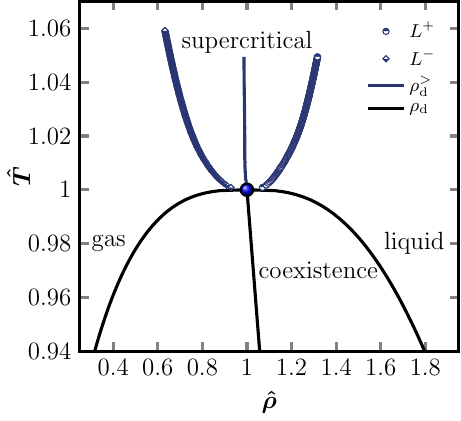}
  \caption{{\bf Supercritical-subcritical correspondence.} The following correspondence is visualized in the $\hat{T}$-$\hat{\rho}$ phase diagram: $L^\pm$ lines vs coexistence curve; supercritical diameter $\rho_{\rm d}^>=\frac{1}{2}(\rho^+ + \rho^-)$ vs subcritical diameter $\rho_{\rm d}=\frac{1}{2}(\rho_{\rm L} + \rho_{\rm G})$. 
The lines are drawn based on the NIST data.
  }
  \label{fig:T_rho}
\end{figure}

All existing studies  focus on asymmetric effects  in the subcritical regime ($T<T_{\rm c}$).
In this work, we switch the attention to the supercritical regime ($T>T_{\rm c}$).
There is no coexistence region in the phase diagram above $T_{\rm c}$, and therefore one cannot define a  diameter of the coexistence curve. Here, we propose a {\it supercritical–subcritical correspondence} grounded in the recently introduced supercritical crossover lines $L^\pm$ (see Fig.~\ref{fig:T_rho})~\cite{li2023thermodynamic, lv2024quantum, wang2024quantum, wang2025analogous, cui2025emergent}. These two lines delineate the boundaries of the supercritical region where liquid and gas become indistinguishable,  
 mirroring 
the subcritical coexistence region. Both the supercritical and coexistence regions represent transitional states between liquid and gas. In particular, a recent study reveals that properties of the supercritical fluid—such as the radial distribution function and transport coefficients—evolve from gas-like behavior near the $L^{-}$ line to liquid-like behavior near the $L^{+}$ line~\cite{jin2025supercritical}. These findings motivate us to view the $L^\pm$ lines as the supercritical counterpart to the subcritical coexistence curve, forming the conceptual basis of the present work. Despite this correspondence, there are essential differences: the subcritical coexistence curve marks a first-order phase transition, whereas the $L^\pm$ lines correspond to supercritical crossovers. It is also worth noting that alternative proposals, including the Widom line~\cite{xu2005relation, luo2014behavior} and the Frenkel line~\cite{brazhkin2012two}, each identify only a single supercritical crossover line. As such, they are insufficient for the present purpose, since analyzing asymmetry inherently requires two distinct boundaries.

Here we employ a linear scaling theory to examine the asymmetric effects, caused by linear mixing of fields, on the behavior of the $L^\pm$ lines near $T_{\rm c}$.
Interestingly, 
we find that the {\it ``supercritical diameter''} $\rho_{\rm d}^> = \frac{1}{2}(\rho^+ + \rho^-)$ violates the rectilinear law and follows  Eq.~(\ref{eq:diameter2}), because the coefficients of correction terms are ``antisymmetric''. 
This surprising coincidence 
rationalizes the proposed subcritical-supercritical correspondence. 
Compared to its subcritical counterpart,  $\rho_{\rm d}^>$ has a practical advantage: its scaling regime is significantly larger, making experimental verification easier.  
This advantage allows us to use the
available data of supercritical fluids  in the 
National Institute of Standards and Technology (NIST) database~\cite{NIST} to verify Eq.~(\ref{eq:diameter2}). The same asymmetric scaling  is further validated by a model of liquid-liquid phase transition, and data extracted from higher-order cumulants of the order parameter distribution.
Our results thus suggest that common scaling laws like Eq.~(\ref{eq:diameter2})  capture  universally the asymmetric effects both below and above  a critical point.

{\bf General results from a linear scaling theory.} 
The thermodynamics near a critical point is described by a universal scaling function~\cite{widom1974critical},
\beq
h_3(h_1, h_2) \simeq  |h_2|^{2-\alpha} f(h_1/|h_2|^{\beta + \gamma}),
\label{eq:scaling function}
\eeq
where $h_1, h_2, h_3$ are respectively the ordering field, thermal field (temperature), and singular part of the  thermodynamic potential. The {\it linear scaling
theory}~\cite{schofield1969parametric,schofield1969correlation}  parameterizes   $h_1$  and $h_2$  with ``polar'' variables $r$ and $\theta$,
$h_1  = ar^{\beta+\gamma}\theta(1-\theta^2)$ and $
h_2  = r(1-b^2\theta^2)$, and assumes a linear function of the order parameter (hence called linear scaling theory), $
\phi_1 = \left(\frac{\partial h_3}{\partial h_1}\right)_{h_2} = k r^\beta \theta$, where $a, k$ are system-dependent fitting parameters, and $b^2 = (\delta-3)/(\delta-1)(1-2\beta)$. Here $\alpha$, $\beta$, $\gamma$ and  $\delta$ are standard critical exponents (see details in SM Sec. S1~\cite{supplement}), which satisfy $\alpha + 2\beta+\gamma = 2$ and $\delta = 1+ \frac{\gamma}{\beta}$.   This linear scaling theory yields universal critical amplitude ratios~\cite{brezin1974universal, fisher1998shape}, but the function $f(x)$ in Eq.~(\ref{eq:scaling function}) depends on the system through the parameters $a$ and $k$.

Following the idea of  {\it complete scaling}~\cite{fisher2000yang, kim2003asymmetric}, we assume that the physical fields $T$, $P$ and the chemical potential $\mu$ in a liquid-gas system are linear mixing of $h_1$, $h_2$ and $h_3$,
\beq
\begin{aligned}
h_{1} &= \Delta\hat{\mu} - a_{3} \tan\varphi \Delta\hat{T} + a_{3} \Delta\hat{P}, \\
h_{2} &= \Delta\hat{T} + b_{2} \Delta\hat{\mu} + \tan\varphi \Delta\hat{P}, \\
h_{3} &= \Delta\hat{P} - \Delta\hat{\mu} - \tan\varphi \Delta\hat{T}.
\label{eq:complete_scalingh123}
\end{aligned}
\eeq
Here  
$\Delta\hat{\mu} \equiv (\mu-\mu_{\rm c})/k_{B}T_{\rm c}$ and $\Delta\hat{P} \equiv (P - P_{\rm c}) /(\rho_c k_B T_c)$. 

It has been suggested that the asymmetric terms  originate from  linear mixing: the coefficients $D_1$ and $D_2$ in Eq.~(\ref{eq:diameter2})  depend on $a_3, b_2$ and $\tan\varphi$ in Eq.~(\ref{eq:complete_scalingh123})~\cite{anisimov1998nature}. Essentially, this linear mixing maps the original order parameter $\phi_1$, which has $Z_2$ symmetry, to a combination of the physical order parameter (density $\rho$) and entropy $S$ (see SM~Sec.~S1~\cite{supplement}). Reverting this mapping then gives rise to the asymmetric behavior of $\rho$.
Note that, the linear mixing Eq.~(\ref{eq:complete_scalingh123}) is independent of the specific choice of parameterization used in the linear scaling theory.

The above setting  provides a standard theoretical model to compute $\rho(P,T)$ and $\kappa_T(P,T)$
near 
the critical point, based on which we can obtain  $L^{\pm}$ (see Appendix A).
The basic idea is to locate the maxima of response functions along lines  in parallel to
the reference line (i.e., the critical isochore $\rho = \rho_{\rm c}$ line). 
Along $L^{\pm}$ lines, $P$ and $\rho$ follow universal scalings when $\Delta \hat{T} \to 0$~\cite{li2023thermodynamic, lv2024quantum, wang2024quantum, wang2025analogous, cui2025emergent},
\beq
\begin{split}
   \delta \hat{P}^{\pm}  &= A_{P}^{\pm}  \Delta \hat{T}^{\Delta},\\
   \delta \hat{\rho}^{\pm}  &= A_{\rho}^{\pm}  \Delta \hat{T}^{\beta},
  \end{split}
    \label{eq:scaling_LG}
   \eeq
where  $\Delta =  \beta + \gamma $ is the gap exponent. Here $\delta \hat{\rho}^{\pm}$ (and $\delta \hat{P}^{\pm}$) quantifies the  difference between   $\rho$ (and  $P$) on the $L^\pm$ lines and that on the critical isochore line (see Fig.~\ref{fig:Z2_illustration}B).

 While the scalings in Eq.~(\ref{eq:scaling_LG}) have been well established in various transitions in liquid-gas~\cite{li2023thermodynamic}, liquid-liquid~\cite{wang2025analogous}, quantum~\cite{lv2024quantum, wang2024quantum}, and black-hole systems~\cite{wang2025analogous, cui2025emergent}, they are only valid to the lowest order. In order to reveal the asymmetric effects, we need to include higher-order scaling corrections~\cite{wegner1976critical}. Specifically, we find that  $A_{P}^{\pm}$ and $A_{\rho}^{\pm}$ in Eq.~(\ref{eq:scaling_LG}) should include additional $T$-dependent terms
 (see SM Sec.~S1~\cite{supplement}),
 \beq
\begin{split}
A_{P}^{\pm}(\Delta \hat{T}) &= A_P^{0,\pm} + A_P^{1,\pm} \Delta \hat{T}^{\Delta-1} + A_P^{2,\pm} \Delta \hat{T}^{1-\alpha} + \cdots,\\
A_{\rho}^{\pm}(\Delta \hat{T}) &= 
A_\rho^{0,\pm} + A_\rho^{1,\pm} \Delta \hat{T}^{\beta} + A_\rho^{2,\pm}  \Delta \hat{T}^{\Delta-1} + \cdots\\
\end{split}
\label{eq:prefactor_form1}
\eeq
 
For the 3D Ising universality class, 
$\beta \approx 0.33$, 
$\Delta - 1 \approx 0.56$ and $1-\alpha \approx 0.89$~\cite{PhysRevE.65.066127,guida1998critical,Chang_2025}. The coefficients satisfy,
\begin{equation}
\begin{split}
&A_P^{0,+}/A_P^{0,-} = - 1,\, A_P^{1,+}/A_P^{1,-}= 1,\, A_P^{2,+}/A_P^{2,-} = -1,\\
&A_\rho^{0,+}/A_\rho^{0,-} = - 1,\, A_\rho^{1,+}/A_\rho^{1,-} = 1,\, A_\rho^{2,+}/A_\rho^{2,-} = 1.
\end{split}
\label{eq:coefficients}
\end{equation}
Very interestingly, the coefficients  of some correction terms are ``antisymmetric'', with a ratio one. These terms are the origin of asymmetric effects, because they will appear in the scaling of the mean (diameter) of $L^\pm$ lines. In contrast, in the Ising model, $A_P^{n,+}/A_P^{n,-} = A_\rho^{n,+}/A_\rho^{n,-} = -1$ for any order $n$, and thus all scaling terms cancel in the expression of the diameter. Note that, even for asymmetric systems, the coefficients of the leading-order term are always symmetric: $A_P^{0,+}/A_P^{0,-} = A_\rho^{0,+}/A_\rho^{0,-} = -1$, consistent with the results in Ref.~\cite{cui2025emergent}.

With details provided in SM Sec.~S1~\cite{supplement}, we outline here the main derivation steps leading to our key results, Eqs.~(\ref{eq:prefactor_form1}) and~(\ref{eq:coefficients}).
(i) Derive the EOSs for the order parameter $\hat{\rho}(r,\theta)$ and susceptibility $\hat{\kappa}_T(r,\theta)$ based on the complete scaling hypothesis, Eq.~(\ref{eq:complete_scalingh123}).
(ii) Determine the $L^{\pm}$ lines, along which the angular coordinates satisfy, to the leading order, a  symmetric relation: $\theta^+ = - \theta^- = \text{constant}$.
(iii) Expand $\Delta \hat{T}$ and $\Delta \hat{\rho}$ in powers of $\delta \hat{P}$ and invert these relations to obtain Eq.~(\ref{eq:prefactor_form1}) with the coefficients $A_{P}^{n,\pm}$ and $A_{\rho}^{n,\pm}$, whose explicit expressions are given in Eqs.~(S29)-(S30).
(iv) Derive the symmetric and asymmetric properties in Eq.~(\ref{eq:coefficients}) from these expressions  (particularly their dependence on $\theta^\pm$) and the  relation $\theta^+ = -\theta^-$.

According to the definitions, $P_{\rm d}^> = 
\frac{1}{2}(P^+ + P^-)$ and $\rho_{\rm d}^> = 
\frac{1}{2}(\rho^+ + \rho^-)$, Eqs.~(\ref{eq:scaling_LG}-\ref{eq:coefficients})  give (with higher-order terms neglected), 
\beq
\begin{split}
P_{\rm d}^> &=  P_{\rm c} + d_P |\Delta \hat{T}|^{2\Delta-1}, \\
\rho_{\rm d}^> &=  \rho_{\rm c} + d_\rho |\Delta \hat{T}|^{2\beta}
+ e_\rho |\Delta \hat{T}|^{1-\alpha},
\label{eq:diameter_super}
\end{split}
\eeq
where $ 
d_P = (A_P^{1,+}+A_P^{1,-})/2$, $d_\rho = (A_\rho^{1,+}+A_\rho^{1,-})/2$ and $e_\rho = (A_\rho^{2,+}+A_\rho^{2,-})/2$.
Note that the expression for $\rho_{\rm d}^>$ is exactly the same as that for $\rho_{\rm d}$ given in Eq.~(\ref{eq:diameter2}).
In contrast, the pressure part  does not have a subcritical counterpart, because the coexistence region shrinks into a single coexistence line in the $P$-$T$ phase diagram (see Fig.~\ref{fig:Z2_illustration}B). The second correction term does not appear in the pressure scaling because of the symmetric 
coefficients, $A_P^{2,+}/A_P^{2,-} = -1$. As shown in SM Sec.~S1.D~\cite{supplement}, the two antisymmetric corrections in the supercritical diameter can be traced to distinct sources of field mixing. The $|\Delta \hat{T}|^{2\beta}$ term originates from the coupling between $\Delta\hat{\mu}$ and $\Delta\hat{P}$, which introduces a nonlinear correction $\propto \phi_1^2$ to the density. 
The $|\Delta \hat{T}|^{1-\alpha}$ term arises from the tilt of the coexistence curve, which adds 
a linear correction $\propto \phi_2$ to the density.
This interpretation is consistent with the subcritical analysis~\cite{PhysRevE.75.051107}.

In order to test the predictions by experimental data, it is useful to combine Eqs.~(\ref{eq:scaling_LG}) and (\ref{eq:prefactor_form1}), and rewrite them  using $|\delta \hat{P}|$ as the independent (control) parameter: 
\beq
\begin{split}
\frac{\Delta \hat{T}^-}{\Delta \hat{T}^+} &= 1 + k_P  |\delta \hat{P}|^{1 - \frac{1}{\Delta}},\\
\frac{\delta \hat{\rho}^-}{\delta \hat{\rho}^+} & = 1 + k_{\rho} |\delta \hat{P}|^{\frac{\beta}{\Delta}}+ m_{\rho} |\delta \hat{P}|^{1 - \frac{1}{\Delta}},
  \end{split}
    \label{eq:ratio}
   \eeq
where $\beta/\Delta \approx 0.21$, $1 - \frac{1}{\Delta} \approx 0.36$ and $\frac{1-\alpha}{\Delta} = 0.57$. It suggests that, for the same given $|\delta \hat{P}|$, the ratio of the values of $\Delta \hat{T}$ and $\delta \hat{\rho}$ on the $L^\pm$ lines should follow Eq.~(\ref{eq:ratio}).

The coefficients $d_P$, $d_\rho$, $e_\rho$, $k_P$, $k_\rho$ and $m_\rho$ in Eqs.~(\ref{eq:diameter_super}) and~(\ref{eq:ratio}) depend on the mixing coefficients $a_3$, $b_2$ and $\tan \varphi$ in Eq.~(\ref{eq:complete_scalingh123}) (their expressions are  given explicitly in SM Sec.~S1~\cite{supplement}).
Importantly, setting  $a_3 = b_2 = \tan \varphi = 0$ causes all of these coefficients to vanish.
Thus, all asymmetric correction terms  come from linear mixing, and are absent in symmetric systems where $h_1 = \Delta \hat{\mu}$ and $h_2 = \Delta \hat{T}$ (such as the lattice gas model). Correspondingly, without linear mixing,the supercritical diameter is rectilinear $\rho_{\rm d}^> \propto |\Delta \hat{T}|$, a contribution from the non-singular, background part of the thermodynamic potential.

The above results, Eqs.~(\ref{eq:prefactor_form1}-\ref{eq:ratio}), are obtained based on linear mixing of three fields, Eq.~(\ref{eq:complete_scalingh123}). If two fields are mixed, e.g., $h_{1} =   -\tan\varphi \Delta\hat{T} +  \Delta\hat{P}$ and $
h_{2} = \Delta\hat{T} + \tan\varphi \Delta\hat{P}$~\cite{luo2014behavior}, then only one asymmetric correction term survives in each expression
(see SM Sec.~S1~\cite{supplement} for details). 

\begin{figure}[!htbp]
  \centering
\includegraphics[width=0.9\linewidth]{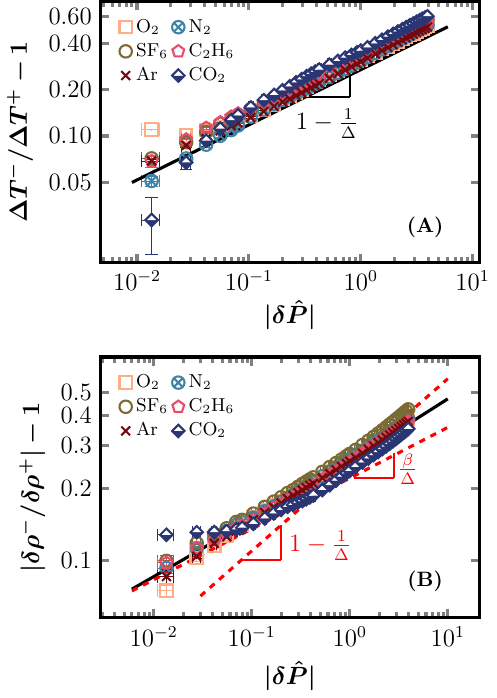}
  \caption{{\bf Asymmetric effects in supercritical fluids}.
  The black solid lines are obtained from fitting with exponents fixed: (A) $y=0.27 x^{0.36}$; (B) $y=0.191 x^{0.21} + 0.07 x^{0.36}$.
  The red dashed lines in (B) represent the two terms in Eq.~(\ref{eq:ratio}).
  }
  \label{fig:NIST}
\end{figure}

{\bf Examination with data of liquid-gas and liquid-liquid phase transitions.}
To test the above theoretical predictions, we  analyze the NIST data of liquid-gas transitions in the following systems~\cite{NIST}: {oxygen ($\rm{O_2}$), nitrogen ($\rm{N_2}$), sulfur hexafluoride ($\rm{SF_6}$), ethane ($\rm{C_2H_6}$), argon ($\rm{Ar}$) and carbon dioxide ($\rm CO_2$).} 
We will examine Eq.~(\ref{eq:ratio}), which
is practically convenient  for an accurate analysis of the data because the $L^\pm$ lines are defined along constant-$\delta \hat{P}$ lines. 
It should be emphasized  that Eq.~(\ref{eq:ratio}) is mathematically equivalent to the more conventional forms Eqs.~(\ref{eq:prefactor_form1}) and~(\ref{eq:diameter_super}).

The $L^\pm$ lines for the above systems are obtained, and the ratios, $\Delta \hat{T}^-/\Delta \hat{T}^+$ and  $\delta \hat{\rho}^-/\delta \hat{\rho}^+$, at the same $\delta \hat{P}$, are plotted in Fig.~\ref{fig:NIST}. 
Interestingly, the data nearly collapse. 
Fitting the data to Eq.~(\ref{eq:ratio}), with the known 3D Ising exponents, gives, $k_P =  0.27 $,  $k_\rho = 0.191$ and $m_\rho = 0.07$. 
The NIST data show remarkable agreement with our theoretical prediction: the pressure scaling (Fig.~\ref{fig:NIST}A) only needs one correction term because only $A_P^{1,\pm}$ are antisymmetric; in contrast, the density scaling (Fig.~\ref{fig:NIST}B) requires two correction terms because both $A_\rho^{1,\pm}$ and $A_\rho^{2,\pm}$ are antisymmetric (see Eq.~\ref{eq:coefficients}).

Our results suggest that all coefficients in Eqs.~(\ref{eq:diameter_super}) and (\ref{eq:ratio})
are universal for the six systems included in Fig.~\ref{fig:NIST}, in contrast to the subcritical case with system-dependent $D_1$ and $D_2$~\cite{anisimov2006nature}.  
However, we do find deviation from this universal behavior for the NIST data of hydrogen $H_2$ (see Appendix B). We also note that this apparent universality is observed given the precision of the NIST data and within the limited temperature range considered ($|\Delta \hat{T}| \in [0.008, 0.3]$). This window is chosen to avoid the immediate vicinity of the critical point, where experimental precision is insufficient, as well as the region far from the critical point, where regular background corrections become non-negligible. 

In general, the subcritical scaling regime of asymmetric corrections appears
to be smaller than the supercritical  scaling regime. For instance, using the NIST data within the same temperature range $|\Delta \hat{T}|\in[0.008,0.3]$ as in Fig.~\ref{fig:NIST}, we find that the subcritical diameter $\rho_{\rm d}$ closely follows the rectilinear law, Eq.~(\ref{eq:diameter2}) (see Appendix C). Thus, the violation of the rectilinear diameter law and therefore the asymmetric effects are more difficult to observe below the critical point.

To examine the universality of the asymmetric scaling, we additionally verify Eq.~(\ref{eq:ratio}) by a mean-field two-state theoretical model of the liquid-liquid phase transition~\cite{bertrand2011peculiar, holten2012entropy}.  The model uses mean-field exponents ($\beta=1/2, \gamma=1, \alpha=0$), with other model parameters obtained in ~\cite{bertrand2011peculiar, holten2012entropy} by fitting the theoretical EOSs to the experimental data of supercooled water (see Appendix D). We find  that only first-order corrections are needed,  $\frac{\Delta \hat{T}^-}{\Delta \hat{T}^+} = 1 + k_P  |\delta \hat{P}|^{1/3}$, and $
\frac{\delta \hat{\rho}^-}{\delta \hat{\rho}^+}  = 1 + k_\rho |\delta \hat{P}|^{1/3}$, consistent with $\beta/\Delta = 1-1/\Delta = 1/3$ for the mean-field model.

\begin{figure}[!htbp]
  \centering
  \includegraphics[width=\linewidth]{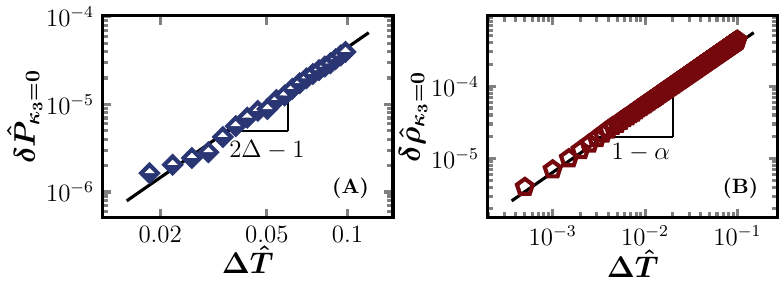}
  \caption{{\bf Asymmetric effects for the ``symmetry line'' with $\kappa_3 = 0$.
  } The black solid lines are obtained from fitting with exponents fixed: (a) $y=0.006x^{2.127}$; (b) $y=0.003x^{0.89}$.
  }
  \label{fig:rho_d}
\end{figure}

{\bf Asymmetry revealed by higher-order cumulants of the order parameter distribution.}
Research interest is developing on  understanding the nature of the supercritical state above the conjectured  critical point in the  quantum chromodynamics (QCD) phase diagram~\cite{sordi2024introducing, shahkarami2025schwinger, hanada2023partial, Glozman2022, fujimoto2025new}, for which experimental exploration relies on the measurement of higher-order moments of fluctuating observables, such as the net-proton multiplicity~\cite{stephanov2011sign, aggarwal2010higher, luo2022properties}. It motivates us to  investigate the asymmetric effects in higher cumulants of the order parameter distribution, in the supercritical regime.

Below we consider the EOSs given by a linear scaling theory with parameters $a=k=0.1$ and {$\varphi = 30^{\circ}$, using  3D Ising universality exponents. To be more explicit, we mix  two fields and therefore only one asymmetric correction term appears in Eq.~(\ref{eq:diameter_super}) ($d_\rho = 0$). 
The $n$-th cumulants $\kappa_n$ of the number distribution in a  grand canonical ensemble are computed using the formulas provided by the fluctuation solution theory~\cite{ploetz2017fluctuation, ploetz2019gas} (see SM Sec.~S2~\cite{supplement}).

The curve with a zero skewness $\kappa_3 \equiv \langle (N- \langle N \rangle) ^{3} \rangle/V = 0$ is called a ``symmetry line'' in Ref.~\cite{ploetz2019gas} for the following reason: the sign of $\kappa_3$ indicates favor of particle insertion (in a gas-like state) or removal (in a liquid-like state), and thus a zero  $\kappa_3$ should correspond to the particle-hole symmetry. However, careful examination of the $\kappa_3 = 0$ ``symmetry line'' shows that it in fact follows the asymmetric scaling, $P_{\kappa_3 = 0} - P_{\rho=\rho_{\rm c}} \sim |\Delta \hat{T}|^{2\Delta-1}$ and $\rho_{\kappa_3 = 0} - \rho_{\rm c} \sim |\Delta \hat{T}|^{1-\alpha}$, as in Eq.~(\ref{eq:diameter_super}) (see Fig.~\ref{fig:rho_d}). 
Thus, even at the loci where the highest symmetry was suggested, the effects of asymmetry nevertheless endure as an intrinsic  property.
In addition, these universal scalings also appear in the lines defined by the maxima of $\kappa_4$ (see SM Sec.~S2~\cite{supplement}).

{\bf Conclusion and Discussion.}
By drawing an analogy between the $L^{\pm}$ lines and the subcritical coexistence curve, we establish a supercritical-subcritical correspondence. This framework leads to a theoretical prediction: asymmetry induces universal scaling corrections with antisymmetric coefficients—a result that aligns remarkably well with the data from the NIST liquid-gas database.

The supercritical-subcritical correspondence could inspire many future research directions. For instance, it would be interesting to compute a supercritical analog  of the latent heat defined by $L^{\pm}$ lines, and examine its relation  to the anomalous phenomenon of supercritical pseudo-boiling~\cite{banuti2015crossing, he2023distinguishing, wang2021three}. 
The  asymmetric effects can be further discussed near the critical points in black hole models~\cite{wang2025analogous, li2025thermodynamic, li2025critical},  QCD systems~\cite{sordi2024introducing} in the framework of
Yang-Mills theory, and quantum magnetic  materials~\cite{jimenez2021quantum, wang2023plaquette}. A rigorous mathematical formulation based on the Lee-Yang theory would also be a compelling avenue to pursue~\cite{ouyang2024complex}.\\

{\bf Acknowledgments.}
We thank Matteo Baggioli, Wei Li, Li Li, and Limei Xu for useful discussions. 
Y. J. acknowledges funding from National Key R\&D Program of China (Grant No. 2025YFF0512000), the China Manned Space Program (Grant No. CMSS-2025-5-P-002), Wenzhou Institute (No. WIUCASICTP2022) and the National Natural Science Foundation of China (No. 12447101).  X. L.  acknowledges the Postdoctoral Fellowship Program of CPSF (Grant Number GZC20252776). We acknowledge the use of the High Performance Cluster at Institute of Theoretical Physics, Chinese Academy of Sciences and the computer clusters at the Hefei advanced computing center.

\bibliography{Z2}

\clearpage

\centerline{\bf \large End Matter}
\vspace{0.5cm}

{\bf Appendix A: the definition of $L^\pm$ lines.}
The $L^\pm$ lines are determined according to the following procedure~\cite{li2023thermodynamic}: (i) define a reference line in the field-temperature phase diagram; (ii) draw a series of paths in parallel to the reference line and find the susceptibility (response of the order parameter to the external field) maximum along each path; (iii) connect the loci of susceptibility maxima in the phase diagram. An example of $L^{\pm}$ is shown in Fig.~\ref{fig:Z2_illustration}b (in the $P$-$T$ phase diagram) and Fig.~\ref{fig:T_rho} (in the $T$-$\rho$ phase diagram). The  reference line is essential in the definition of $L^{\pm}$. For the Ising model, the choice is trivial -- the reference line is the $T$-axis, which coincides with the coexistence line (see Fig.~\ref{fig:Z2_illustration}a). For the liquid-gas and other asymmetric systems, a natural choice is the  critical isochore, i.e., the line with a constant critical density (order parameter) $\rho_{\rm c}$. Under the above definition, $\delta \hat{P}^{\pm}$  quantifies the distance ($P$ difference) from the reference line to $L^{\pm}$ at the given $T$, and  $\delta \hat{\rho}^{\pm}$ quantifies their density difference. 
Note that we use notations with ``$\delta$'' (e.g., $\delta {P} = P(\rho,T) - P(\rho_{\rm c},T)$) to denote  the distance to the critical isochore, and those with ``$\Delta$'' (e.g., $\Delta P = P - P_{\rm c}$) to denote the distance to the critical point. Note that $\delta \hat{\rho}^{\pm} = \Delta \hat{\rho}^{\pm}$.\\

{\bf Appendix B: asymmetric effects in supercritical hydrogen.}
Fig.~\ref{fig:H2} shows that the supercritical hydrogen data obtained from 
NIST deviate from the behavior of systems listed in Fig.~\ref{fig:NIST}. The deviation of hydrogen from the universal scaling observed in Fig.~3 may originate from two sources. First, experimental data near the critical point are often less accurate for some substances, and such inaccuracies can directly affect the scaling behavior extracted from the data. Second, the empirical EOSs in the NIST database are typically fitted piecewise over different thermodynamic ranges; the matching between adjacent fitting intervals may introduce weak discontinuities in the EOS or its derivatives. If the range used to extract the $L^{\pm}$ lines straddles the boundaries of these intervals, deviations from the  universal scaling can arise. A detailed analysis of these effects is beyond the scope of the present work.\\

\begin{figure}[!htbp]
  \centering
  \includegraphics[width=\linewidth]{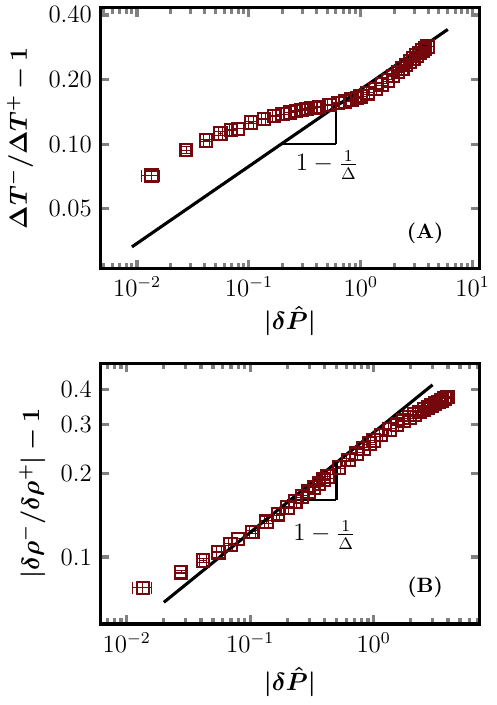}
  \caption{{\bf Asymmetric effects in supercritical hydrogen.} The solid fitting line in (A) represents $y=0.18 x^{0.36}$, and the one in (B) represents $y= 0.28 x^{0.36}$.
  }
  \label{fig:H2}
\end{figure}

{\bf Appendix C: subcritical diameter of carbon dioxide.}
Fig.~\ref{fig:subrhoD} shows that the subcritical diameter $\hat{\rho}_{\rm d}$ of carbon dioxide follows closely the rectilinear law, Eq.~(\ref{eq:diameter}) - the asymmetric corrections are nearly unobservable. Note that the data  plotted in Fig.~(\ref{fig:NIST}) (supercritical) and those in Fig.~\ref{fig:subrhoD} (subcritical) cover approximately the same temperature range ${\Delta \hat{T}} \in [0.008, 0.3]$. 
\\

\begin{figure}[!htbp]
  \centering
  \includegraphics[width=\linewidth]{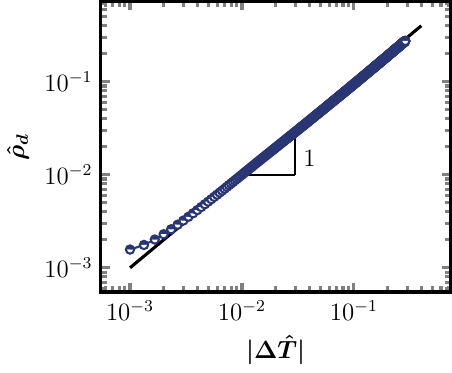}
  \caption{{\bf Asymmetric effects in subcritical carbon dioxide.} The solid  line is $y=x$.
  }
  \label{fig:subrhoD}
\end{figure}

\begin{figure}[!htbp]
  \centering
  \includegraphics[width=\linewidth]{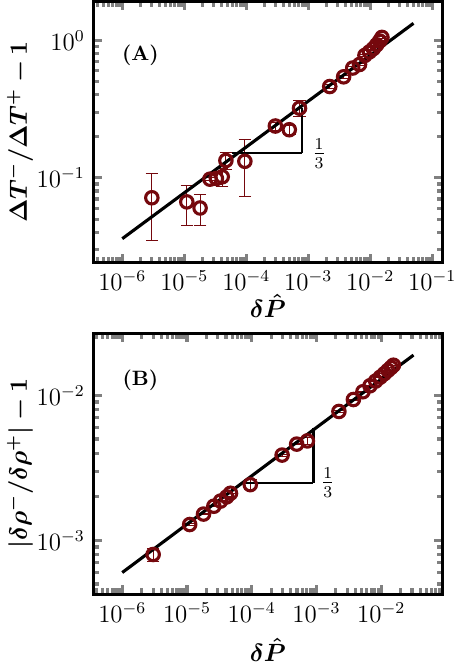}
  \caption{{\bf Asymmetric effects in $L^\pm$ lines examined by the two-state model of the liquid-liquid phase transitions.} The black solid lines are obtained from fitting with exponents fixed: (a) $y=3.6x^{\frac{1}{3}}$; (b) $y=0.06x^{\frac{1}{3}}$.
  Data are consistent with Eq.~(\ref{eq:ratio}) with mean-field exponents.
  }
  \label{fig:water}
\end{figure}

{\bf Appendix D: two-state model for liquid-liquid transitions.}
The two-state model offers a thermodynamic framework for describing polyamorphic single-component liquids by treating them as a binary mixture of two distinct but interconvertible liquid states, denoted as A and B, with corresponding concentrations \(1-x\) and \(x\). 
The present analysis adopts the mean-field formulation of the two-state model, as detailed in Refs.~\cite{bertrand2011peculiar, holten2012entropy}.
The molar Gibbs free energy for this binary mixture is,
\begin{equation}
\frac{G}{k_{\mathrm{B}}T} = \frac{G^{\mathrm{A}}}{k_{\mathrm{B}}T} + x\frac{G^{\mathrm{BA}}}{k_{\mathrm{B}}T} + x\ln x + (1 - x)\ln(1 - x) + \omega x(1 - x),
\label{eq:G}
\end{equation}
where \(k_{\mathrm{B}}\) is the Boltzmann constant. The term \(x\ln x + (1 - x)\ln(1 - x)\) corresponds to the ideal entropy of mixing, while \(\omega x(1 - x)\) represents the excess mixing entropy arising from interactions between the A and B states. The interaction parameter is defined as \(\omega = 2 + \omega_0\Delta P\), where \(\omega_0\) is a fitting parameter.
The quantity \(G^{\mathrm{A}} = \sum_{m,n} c_{mn} (\Delta T)^{m} (\Delta P)^{n}\) represents the Gibbs free energy of the pure A state, determined through fitting experimental data with adjustable coefficients \(c_{mn}\). The free energy difference between states B and A is given by \(G^{\mathrm{BA}} = G^{\mathrm{B}} - G^{\mathrm{A}}\). In the vicinity of liquid-liquid phase transitions, this difference is approximated by:
\begin{equation}
G^{\mathrm{BA}}/k_{\mathrm{B}}T = \lambda(\Delta T + \alpha_1\Delta P + \alpha_2\Delta T\Delta P),
\label{eq:GBA}
\end{equation}
where \(\lambda\), \(\alpha_1\), and \(\alpha_2\) are additional fitting parameters. 
The condition \(G^{\mathrm{BA}} = 0\) defines the coexistence line, with parameters \(\alpha_1\) and \(\alpha_2\) characterizing its slope and curvature, respectively.

The equilibrium fraction \(x\) is obtained by solving:
\begin{equation}
\begin{split}
\mu^{\mathrm{BA}} 
= G^{\mathrm{BA}} + k_{\mathrm{B}}T\left[\ln\frac{x}{1 - x} + \omega(1 - 2x)\right] = 0,
\label{eq:mu}
\end{split}
\end{equation}
where $\mu^{\mathrm{BA}} = \left(\frac{\partial G}{\partial x}\right)_{T,P}$.
The solution of $x$ is then  substituted 
into Eq.~(\ref{eq:G}). The critical parameters \(T_{\mathrm{c}}\), \(\rho_{\mathrm{c}}\), and \(P_{\mathrm{c}}\) are also fitting variables. This model exhibits critical scaling behavior consistent with mean-field exponents. Based on these expressions, EOSs can be derived. For instance, the reduced density and reduced volume are given by:
$$
\frac{1}{\hat{\rho}(\hat P, \hat T)} = \hat{V}(\hat P, \hat T) = \left(\frac{\partial \hat{G}}{\partial \hat{P}}\right)_{\hat T},
$$
employing Eq.~(\ref{eq:G}). 

The EOSs of this model are  fitted to experimental data of supercooled water in Ref.~\cite{holten2012entropy},  spanning temperatures from 140 K to 310 K and pressures from 0.1 MPa to 400 Mpa. The optimal fit yields critical parameters {$T_{\rm c} = 227.42~\rm{K}$, $P_{\rm c}=13.45~\rm{MPa}$}. Additional fitting parameters are provided in~\cite{holten2012entropy}. Note that we choose these fitting parameters to make the model concrete. The scaling behavior is independent of these fitting parameters.

The \(L^{\pm}\) loci are identified by determining the maxima of the isothermal compressibility \(\kappa_T = \frac{1}{\rho}\left( \frac{\partial \rho}{\partial P}\right)_T \)  
following the procedure described in Appendix A. Fig.~\ref{fig:water} shows the obtained results. Similar to liquid-gas transitions, the  data can be fitted to Eq.~(\ref{eq:ratio}). Because $\beta/\Delta = 1 - 1/\Delta = 1/3$ in the mean-field model, the two correction terms for the density part have the same exponents.

\clearpage

\setcounter{figure}{0}
\setcounter{equation}{0}
\setcounter{table}{0}
\setcounter{section}{0}
\renewcommand\thefigure{S\arabic{figure}}
\renewcommand\theequation{S\arabic{equation}}
\renewcommand\thesection{S\arabic{section}}
\renewcommand\thetable{S\arabic{table}}

\centerline{\bf \large Supplementary  Material}
\vspace{1cm}
\tableofcontents
\vspace{1cm}

\section{Scaling theory}

\subsection{Linear scaling theory}

The linear scaling
theory~\cite{schofield1969parametric,schofield1969correlation}  parameterizes   the two external fields $h_1$ (ordering field) and $h_2$ (thermal field) with ``polar'' variables $r$ and $\theta$,
\beq
\begin{aligned}
h_1  &= ar^{\beta+\gamma}\theta(1-\theta^2),\\
h_2  &= r(1-b^2\theta^2),
\label{eq:h_12}
\end{aligned}
\eeq
where $\beta$ and $\delta$ are standard critical exponents,
$a$ is a system-dependent fitting parameter, and $b$ is a universal parameter obeying  $b^2 = (\delta-3)/(\delta-1)(1-2\beta)$. The variable $r$ represents a ``distance'' from the critical point, and $\theta\in[-1,1]$ represents the ``angle'' from the coexistence line ($\theta = \pm 1$).
The critical part of themodynamic potential of a system near the critical point is a homogeneous function of $h_1$ and $h_2$,
\beq
h_3=\left|h_{2}\right|^{2-\alpha} f\left(h_{1} /\left|h_{2}\right|^{\beta+\gamma}\right),
\label{eq:homogeneous funtion}
\eeq
parameterized as
\beq
h_3 = r^{2-\alpha}p(\theta),
\label{eq:themodynamic potential}
\eeq
where $p(\theta)$ is an analytical function. The  differential form of $ h_3 $ is given by:
\begin{equation}
    d h_3 = \phi_1  d h_1 + \phi_2  d h_2.
    \label{eq:scaling_form}
\end{equation}
The order parameter conjugated to $h_1$ is,
\beq
\begin{aligned}
\phi_1 = \left(\frac{\partial h_3}{\partial h_1}\right)_{h_2}  
       =r^\beta m(\theta)
         = k r^\beta \theta,
\label{eq:order parameter1}
\end{aligned}
\eeq
where $m (\theta)$ is assumed to be a linear function of $\theta$, $m(\theta) =k\theta$ 
~\cite{schofield1969parametric,schofield1969correlation}, and $k$ is another system-dependent fitting parameter. The thermodynamic  parameter (entropy) conjugated to $h_2$ is,
\beq
\begin{aligned}
\phi_2 = \left(\frac{\partial\Psi}{\partial h_2}\right)_{h_1} 
       =r^{1-\alpha} s(\theta).
\label{eq:order parameter2}
\end{aligned}
\eeq
The function $s(\theta)$ can be determined from the equality $\left(\frac{\partial\phi_1}{\partial h_2}\right)_{h_1}=\left(\frac{\partial\phi_2}{\partial h_1}\right)_{h_2}$, 
\beq
s(\theta) &=a k(s_0+s_2\theta^2), 
\eeq
where,
\beq
\begin{aligned}
s_0 &= -{\frac{\beta\left[-3+\delta+b^2(-1+\delta)(-2+\beta+\beta\delta)\right]}{2b^4(-2+\beta+\beta\delta)(-1+\beta+\beta\delta)}},\\
s_2 &= \frac{\beta(\delta-3)}{2b^2(-2+\beta+\beta\delta)}.
\label{eq:order parameter2 coefficient}
\end{aligned}
\eeq


The susceptibilities can be computed using the following expressions,
\beq
\begin{aligned}
&\chi_1 \equiv 
\left(\frac{\partial\phi_1}{\partial h_1}\right)_{h_2}= \frac{k}{a}r^{-\gamma}C_1(\theta),\\
&\chi_2 \equiv \left(\frac{\partial\phi_2}{\partial h_2}\right)_{h_1}= r^{-\alpha}C_2(\theta),\\
&\chi_{12} \equiv \left(\frac{\partial\phi_1}{\partial h_2}\right)_{h_1}=\left(\frac{\partial\phi_2}{\partial h_1}\right)_{h_2}= k r^{\beta-1}C_{12}(\theta),
\label{eq:Response function}
\end{aligned}
\eeq
where
\beq
\begin{aligned}
&C_1(\theta)=\frac{(1-b^2\theta^2(1-2\beta))}{C_0(\theta)},\\
&C_2(\theta)=\frac{\left[(1-\alpha)(1-3\theta^2)s(\theta)-2s_2\beta\delta\theta^2(1-\theta^2)\right]}{C_0(\theta)},\\
&C_{12}(\theta)=\frac{\beta\theta\left[1-\delta-\theta^2(3-\delta)\right]}{C_0(\theta)},\\
&C_0(\theta)=(1-3\theta^2)(1-b^2\theta^2)+2\beta\delta b^2\theta^2(1-\theta^2).
\label{eq:Response function coefficient}
\end{aligned}
\eeq

In a symmetric model such as the lattice-gas model, $h_1$ and $h_2$ are related to the physical fields through, 
\beq
\begin{aligned}
&h_{1} = \Delta\hat{\mu} = (\mu - \mu_{\rm c})/k_{B}T_{\rm c},\\
&h_{2} = \Delta\hat{T} = (T - T_{\rm c}) / T_{\rm c}.
\end{aligned}
\eeq
The corresponding order parameters are
$\phi_{1} = \Delta\hat{\rho}$ (density change) and $\phi_{2} = \Delta(\hat{\rho}\hat{S})$ (entropy density change). The quantity $-h_{3}$ represents the critical part of the grand thermodynamic potential ($\Omega = -PV$) per unit volume,  given by:
\begin{equation}
h_{3} = (P - P_{\text{coex}})/(\rho_c k_B T_{\rm c}),
\end{equation}
where $P_{\text{coex}}$ denotes the vapor-liquid coexistence pressure, corresponding to the thermodynamic condition $h_{1} = 0$.

\subsection{Linear mixing of three fields}

To incorporate asymmetry, the scaling fields in the linear scaling theory are expressed as linear combinations of the three physical external fields~\cite{fisher2000yang, kim2003asymmetric,PhysRevE.75.051107,anisimov2006nature}:
\beq
\begin{aligned}
h_{1} &= a_{1} \Delta\hat{\mu} + a_{2} \Delta\hat{T} + a_{3} \Delta\hat{P}, \\
h_{2} &= b_{1} \Delta\hat{T} + b_{2} \Delta\hat{\mu} + b_{3} \Delta\hat{P}, \\
h_{3} &= c_{1} \Delta\hat{P} + c_{2} \Delta\hat{\mu} + c_{3} \Delta\hat{T},
\label{eq:complete_scaling}
\end{aligned}
\eeq
where $\Delta\hat{P} = (P - P_{\rm c}) /(\rho_c k_B T_{\rm c})$.
A simplification in defining the coefficients arises from the fact that any two independent critical amplitudes can be included in the homogeneous function $f(x)$ in Eq.~(\ref{eq:homogeneous funtion}); it is therefore convenient to set $a_1 = 1$, $b_1 = 1$ and $c_1 = 1$. Along the coexistence curve and its supercritical extension, defined by $h_1 = 0$, the chemical potential difference $\Delta\hat{\mu}$ vanishes. Thus, the coupling coefficient between $\Delta\hat{P}$ and $\Delta\hat{T}$ is naturally given by the slope of the coexistence curve, $\tan \varphi$, where $\varphi$ is the slope of the coexistence line. Besides, according to the thermodynamic relation
\beq
dP = \rho \, d\mu + \rho S \, dT,
\label{eq:thermodynamic_identity}
\eeq 
the critical entropy density is consequently given by $
\hat{\rho_{\rm c}}\hat{{S}_{\rm c}} = \left( \frac{\partial \hat{P}}{\partial \hat{T}} \right)_{h_{1} = 0, c}
= \tan \varphi$. Then we can rewrite Eq.~(\ref{eq:complete_scaling}) by setting $c_2 = -\rho_{\rm c} = -1$ and $c_3 = -\rho_{\rm c} S_{\rm c} = -\tan \varphi$:
\beq
\begin{aligned}
h_{1} &= \Delta\hat{\mu} - a_{3} \tan\varphi \Delta\hat{T} + a_{3} \Delta\hat{P}, \\
h_{2} &= \Delta\hat{T} + b_{2} \Delta\hat{\mu} + \tan\varphi \Delta\hat{P}, \\
h_{3} &= \Delta\hat{P} - \Delta\hat{\mu} - \tan\varphi \Delta\hat{T}.
\label{eq:complete_scaling2}
\end{aligned}
\eeq
Inverting the expressions gives  the physical external fields, 
\beq
\begin{split}
\Delta\hat{\mu} &= \frac{A_{1}}{K}h_{1}+\frac{B_{1}}{K}h_{2}+\frac{C_{1}}{K}h_{3}, \\
\Delta\hat{T} &= \frac{A_{2}}{K}h_{1}+\frac{B_{2}}{K}h_{2}+\frac{C_{2}}{K}h_{3}, \\
\Delta\hat{P} &= \frac{A_{3}}{K}h_{1}+\frac{B_{3}}{K}h_{2}+\frac{C_{3}}{K}h_{3},
\end{split}
\label{eq:physical_field}
\eeq
where  
\beq
\begin{aligned}
A_{1} &= b_{1}c_{1}-b_{3}c_{3}, \\
B_{1} &= -a_{2}c_{1}+a_{3}c_{3}, \\
C_{1} &= a_{2}b_{3}-a_{3}b_{1},
\end{aligned}
\eeq
\beq
\begin{aligned}
A_{2} &= -b_{2}c_{1} + b_{3}c_{2}, \\
B_{2} &= a_{1}c_{1} - a_{3}c_{2}, \\
C_{2} &= -a_{1}b_{3} + a_{3}b_{2},
\end{aligned}
\eeq
\beq
\begin{aligned}
A_{3} &= b_{2}c_{3} - b_{1}c_{2}, \\
B_{3} &= -a_{1}c_{3} + a_{2}c_{2}, \\
C_{3} &= a_{1}b_{1} - a_{2}b_{2},
\end{aligned}
\eeq
and $K = a_{1}(b_{1}c_{1} - b_{3}c_{3}) - a_{2}(b_{2}c_{1} - b_{3}c_{2}) + a_{3}(b_{2}c_{3} - b_{1}c_{2})$.

Combining Eqs.~(\ref{eq:scaling_form}),~(\ref{eq:thermodynamic_identity}) and (\ref{eq:complete_scaling2}), we obtain the expressions for the order parameters as follows:
\begin{equation}
\begin{aligned}
\hat{\rho} &= \left( \frac{\partial \hat{P}}{\partial \hat{\mu}} \right)_{\hat{T}} = \frac{1 + \phi_{1} + b_{2} \phi_{2}}{1 - a_{3} \phi_{1} - \tan \varphi \phi_{2}} ,\\
\quad \hat{\rho}\hat{S} &= \left( \frac{\partial \hat{P}}{\partial \hat{T}} \right)_{\hat{\mu}} = \frac{\hat{\rho_{\rm c}}\hat{{S}_{\rm c}} - a_{3} \tan\varphi \phi_{1} + \phi_{2}}{1 - a_{3} \phi_{1} - \tan \varphi \phi_{2}}.
\label{eq:combined_definition}
\end{aligned}
\end{equation}
Performing a Taylor expansion of Eq.~(\ref{eq:combined_definition}) 
and truncating beyond the linear term in the scaling variable $r$ leads to the following approximation:
\beq
\begin{aligned}
\Delta\hat{\rho} &\simeq (1 + a_3)\phi_1 + a_3(1 + a_3)\phi_1^2 + (b_{2}+\tan{\varphi})\phi_2 \\
&\simeq r^{\beta} \left[ \left(1+a_{3}\right) k \theta \right.+ a_{3}\left(1+a_{3}\right) k^{2} r^{\beta} \theta^{2} \\
&+ \left. (b_{2}+\tan{\varphi}) r^{\beta+\gamma - 1} s(\theta) \right].
\label{eq:rho}
\end{aligned}
\eeq

The susceptibility is
\beq
\begin{aligned}
\hat{\kappa}_T &= (\partial \hat{\rho}/\partial \hat{\mu})_{\hat{T}}\\
&=  \left(1+a_{3}\right)\left(1+\frac{2 a_{3}}{1+a_{3}} \phi_{1}\right) \chi_{1} \\
                  & +\left[b_{2}+\tan\varphi+ b_{2} \left(1+a_{3}\right)\left(1+\frac{2 a_{3}}{1+a_{3}} \phi_{1}\right)\right] \chi_{12}\\
                  & + b_{2}\left(b_{2}+\tan\varphi\right) \chi_{2}.
\end{aligned}
\label{eq:susceptibility}
\eeq
From Eq.~(\ref{eq:h_12}), we know $r = \left( \frac{h_{1}}{a \theta (1 - \theta^{2})} \right)^{\frac{1}{\beta+\gamma}}$. Substituting this relationship and Eqs.~(\ref{eq:order parameter1}) and (\ref{eq:Response function}) into the expression above yields:
\begin{equation}
\begin{split}
\hat{\kappa}_T(h_{1},\theta) = h_{1}^{-\frac{\gamma}{\beta+\gamma}} \left\{ A + Bh_{1}^{\frac{\beta}{\beta+\gamma}} + Ch_{1}^{\frac{\beta+\gamma-1}{\beta+\gamma}} \right. \\
\left. + Dh_{1}^{\frac{2\beta+\gamma-1}{\beta+\gamma}} + Eh_{1}^{\frac{\gamma-\alpha}{\beta+\gamma}} \right\},
\label{kappa_h1_theta}
\end{split}
\end{equation}
where
\begin{equation}
\begin{aligned}
A &= \frac{k(1+a_{3})}{a} C_{1}(\theta) \cdot \left( a \theta (1-\theta^2) \right)^{\frac{\gamma}{\beta+\gamma}} \\
B &= \frac{2 a_{3}k^{2}}{a} \theta C_{1}(\theta) \cdot \left( a \theta (1-\theta^2) \right)^{\frac{\gamma-\beta}{\beta+\gamma}} \\
C &= \left(\tan\varphi+2b_{2}+a_{3}b_{2}\right)k C_{12}(\theta) \cdot \left( a \theta (1-\theta^2) \right)^{ \frac{1-\beta}{\beta+\gamma}} \\
D &= 2 a_{3}b_{2}k^{2} \theta C_{12}(\theta) \cdot \left( a \theta (1-\theta^2) \right)^{ \frac{1-2\beta}{\beta+\gamma}} \\
E &= b_{2}\left(b_{2}+\tan\varphi\right) C_{2}(\theta) \cdot \left( a \theta (1-\theta^2) \right)^{ \frac{\alpha}{\beta+\gamma}}.
\end{aligned}
\end{equation}

Next, we show that $\theta$ is a constant along the $L^{\pm}$ lines.
In fact, the quantity $\delta\hat{P}$  is  equivalent to $h_1$ since fixing their values corresponds to paths parallel to the critical isochore. When all the higher-order terms are neglected, the solution that satisfies $\partial \hat{\kappa}_T/\partial\theta = 0$ is found to be independent of $h_1$. Consequently, the extremum of $\hat{\kappa}_T$ on such a path, which we define as $L^{\pm}$, corresponds to a constant value of $\theta$. If we substitute $r = \left( \frac{\delta\hat{P}}{a \theta (1 - \theta^{2})} \right)^{\frac{1}{\beta+\gamma}}$ into Eq.~(\ref{eq:physical_field}) and~(\ref{eq:rho}), $L^{\pm}$ lines satisfy
\begin{equation}
\begin{aligned}
\Delta\hat{T}&=\frac{A_{2}}{K}\delta\hat{P}+\frac{B_{2}}{K} \left( \frac{\delta\hat{P}}{a \theta^{\pm} (1 - \theta^{\pm2})} \right)^{\frac{1}{\beta+\gamma}} \left(1-b^{2}\theta^{\pm2}\right)\\
&+\frac{C_{2}}{K} \left( \frac{\delta\hat{P}}{a \theta^{\pm} (1 - \theta^{\pm2})} \right)^{\frac{2-\alpha}{\beta+\gamma}} p(\theta^{\pm})\\
&= \tilde{A}_P \delta \hat{P}^{\frac{1}{\beta+\gamma}} ,
\end{aligned}
\end{equation}
and
\begin{equation}
\begin{aligned}
\Delta\hat{\rho} &\simeq \tilde{A}_{\rho} \delta\hat{P}^{\frac{\beta}{\beta+\gamma}},  
\label{eq:rho_deltaP}
\end{aligned}
\end{equation}
where
\begin{equation}
\begin{aligned}
\tilde{A}_{P}&= \tilde{A}_P^{0,\pm} + \tilde{A}_P^{1,\pm} \delta \hat{P}^{\frac{\beta+\gamma-1}{\beta+\gamma}} + \tilde{A}_P^{2,\pm} \delta \hat{P}^{\frac{1-\alpha}{\beta+\gamma}}\\
&=\frac{B_{2}}{K} \left( a\theta^{\pm}(1-\theta^{\pm2}) \right)^{-\frac{1}{\beta+\gamma}} (1-b^{2}\theta^{\pm2}) \\
&\quad +  \frac{A_{2}}{K} \delta \hat{P}^{\frac{\beta+\gamma-1}{\beta+\gamma}} + \frac{C_{2}}{K} \left( a\theta^{\pm}(1-\theta^{\pm2}) \right)^{-\frac{2-\alpha}{\beta+\gamma}} \delta \hat{P}^{\frac{1-\alpha}{\beta+\gamma}} p(\theta^{\pm}), 
\label{eq:Ap}
\end{aligned}
\end{equation}
and
\begin{equation}
\begin{aligned}
\tilde{A}_{\rho}&= \tilde{A}_{\rho}^{0,\pm} + \tilde{A}_{\rho}^{1,\pm} \delta \hat{P}^{\frac{\beta}{\beta+\gamma}} + \tilde{A}_{\rho}^{2,\pm} \delta \hat{P}^{\frac{\beta+\gamma-1}{\beta+\gamma}}\\
&= (1 + a_{3}) k\theta^{\pm} \left( a\theta^{\pm}(1-\theta^{\pm2}) \right)^{-\frac{\beta}{\beta+\gamma}}\\
&+ a_{3}(1 + a_{3}) k^{2} \theta^{\pm2} \left( a\theta^{\pm}(1-\theta^{\pm2}) \right)^{-\frac{2\beta}{\beta+\gamma}} \delta\hat{P}^{\frac{\beta}{\beta+\gamma}} \\
&+ (b_{2}+\tan{\varphi}) s(\theta^{\pm}) \left( a\theta^{\pm}(1-\theta^{\pm2}) \right)^{-\frac{1-\alpha}{\beta+\gamma}} \delta\hat{P}^{\frac{\beta+\gamma-1}{\beta+\gamma}}. 
\label{eq:Arho}
\end{aligned}
\end{equation}

{Based on the above results, we can explicitly give  the expressions of all coefficients in Eqs.~(\ref{eq:prefactor_form1}-\ref{eq:ratio}) in the main text:} 

\begin{equation}
\begin{aligned}
{A}_P^{0,\pm}& =(\tilde{A}_P^{0,\pm})^{-\Delta}\\
&=a\theta^{\pm}\left(1-\theta^{\pm2}\right)\left(\frac{B_{2}}{K}\left(1-b^{2}\theta^{\pm2}\right)\right)^{-\Delta} \\
{A}_P^{1,\pm}&=-\Delta{\tilde{A}_P^{1,\pm}}(\tilde{A}_P^{0,\pm})^{-2\Delta}\\
&=-\Delta\frac{A_{2}}{K}a^{2}\theta^{\pm2}\left(1-\theta^{\pm2}\right)^{2}\left(\frac{B_{2}}{K}\left(1-b^{2}\theta^{\pm2}\right)\right)^{-2\Delta} \\
{A}_P^{2,\pm}&=-\Delta{\tilde{A}_P^{2,\pm}}(\tilde{A}_P^{0,\pm})^{-\Delta+\alpha-2}\\
&=-\Delta\frac{C_{2}}{K}a\theta^{\pm}\left(1-\theta^{\pm2}\right)p(\theta^{\pm})\left(\frac{B_{2}}{K}\left(1-b^{2}\theta^{\pm2}\right)\right)^{\alpha-\Delta-2},
\label{eq:Ap}
\end{aligned}
\end{equation}

\begin{equation}
\begin{aligned}
{A}_{\rho}^{0,\pm}& =\tilde{A}_{\rho}^{0,\pm}(\tilde{A}_P^{0,\pm})^{-\beta}\\
&= (1 + a_3) k \theta^{\pm} \left( \frac{B_2}{K} \right)^{-\beta} (1 - b^2 \theta^{\pm2})^{-\beta} \\
{A}_{\rho}^{1,\pm}&=\tilde{A}_{\rho}^{1,\pm}(\tilde{A}_P^{0,\pm})^{-2\beta}\\
&= a_{3}(1 + a_{3})k^{2}\theta^{\pm 2} \left( \frac{B_{2}}{K} \right)^{-2\beta} \left(1 - b^{2}\theta^{\pm 2} \right)^{-2\beta}, \\
{A}_{\rho}^{2,\pm}&={\tilde{A}_{\rho}^{2,\pm}}(\tilde{A}_P^{0,\pm})^{\alpha-1},\\
&= (b_{2} + \tan\varphi) s(\theta^{\pm}) \left( \frac{B_{2}}{K} \right)^{\alpha-1} \left(1 - b^{2}\theta^{\pm 2} \right)^{\alpha-1},
\label{eq:Arho}
\end{aligned}
\end{equation}

\beq
\begin{aligned}
d_P &= \frac{1}{2} \left( A_P^{1,+} + A_P^{1,-} \right) \\
    &= -\frac{\Delta A_2 a^2}{2K} \\
    &\times\sum_{\sigma=\pm} (\theta^\sigma)^2 (1-(\theta^\sigma)^2)^2
        \left( \frac{B_2}{K} (1 - b^2 (\theta^\sigma)^2) \right)^{-2\Delta}, \\[8pt]
d_\rho &= \frac{1}{2} \left( A_\rho^{1,+} + A_\rho^{1,-} \right) \\
       &= \frac{a_3(1+a_3)k^2}{2} \left( \frac{B_2}{K} \right)^{-\beta}\\
        &\times\sum_{\sigma=\pm} (\theta^\sigma)^2 \left( a\theta^\sigma (1-(\theta^\sigma)^2) \right)^{-\frac{\beta}{\Delta}}
          (1 - b^2 (\theta^\sigma)^2)^{-\beta}, \\[8pt]
e_\rho &= \frac{1}{2} \left( A_\rho^{2,+} + A_\rho^{2,-} \right) \\
       &= \frac{b_2 + \tan\phi}{2} \left( \frac{B_2}{K} \right)^{-\beta}\\
    &\times\sum_{\sigma=\pm} s(\theta^\sigma) \left( a\theta^\sigma (1-(\theta^\sigma)^2) \right)^{\frac{\alpha+\beta-1}{\Delta}}
          (1 - b^2 (\theta^\sigma)^2)^{-\beta},
\label{eq:d1d2}
\end{aligned}
\eeq

\begin{equation}
\begin{aligned}
k_P&=\tilde{A}_{P}^{1,-}/\tilde{A}_{P}^{0,-}-\tilde{A}_{P}^{1,+}/\tilde{A}_{P}^{0,+}\\
&= \frac{A_{2}}{B_{2}}\left[\frac{\left(a\theta^-\left(1-\theta^{-2}\right)\right)^{\frac{1}{\beta+\gamma}}}{1-b^{2}\theta^{-2}}-\frac{\left(a\theta^+\left(1-\theta^{+2}\right)\right)^{\frac{1}{\beta+\gamma}}}{1-b^{2}\theta^{+2}}\right],
\\
\label{eq:kpmp}
\end{aligned}
\end{equation}

\begin{equation}
\begin{aligned}
k_{\rho} &=\tilde{A}_{\rho}^{1,-}/\tilde{A}_{\rho}^{0,-}-\tilde{A}_{\rho}^{1,+}/\tilde{A}_{\rho}^{0,+}\\
&= a_{3}k\theta^{-}\left(a\theta^{-}(1-\theta^{-2})\right)^{-\frac{\beta}{\beta+\gamma}}\\
&-a_{3}k\theta^{+}\left(a\theta^{+}(1-\theta^{+2})\right)^{-\frac{\beta}{\beta+\gamma}},
\\
m_{\rho} &=\tilde{A}_{\rho}^{2,-}/\tilde{A}_{\rho}^{0,-}-\tilde{A}_{\rho}^{2,+}/\tilde{A}_{\rho}^{0,+}\\&= \frac{\left(b_{2}+\tan\varphi\right)s(\theta^{-})}{\left(1+a_{3}\right)k\theta^{-}}\left(a\theta^{-}(1-\theta^{-2})\right)^{\frac{\alpha+\beta-1}{\beta+\gamma}}\\
&-\frac{\left(b_{2}+\tan\varphi\right)s(\theta^{+})}{\left(1+a_{3}\right)k\theta^{+}}\left(a\theta^{+}(1-\theta^{+2})\right)^{\frac{\alpha+\beta-1}{\beta+\gamma}}.
\label{eq:kpmp}
\end{aligned}
\end{equation}

\vspace{10pt}  

\subsection{Symmetric and antisymmetric corrections}
Now let us consider the symmetry of the terms in the following expansion:

\beq
\begin{split}
A_{P}^{\pm}(\Delta \hat{T}) &= a_{P}^\pm \left(1+b_{P}^{\pm} \Delta \hat{T}^{\Delta-1} + c_P^{\pm} \Delta \hat{T}^{1-\alpha} + \cdots \right) \\
&= A_P^{0,\pm} + A_P^{1,\pm} \Delta \hat{T}^{\Delta-1} + A_P^{2,\pm} \Delta \hat{T}^{1-\alpha} + \cdots,\\
\end{split}
\eeq
and
\beq
\begin{split}
A_{\rho}^{\pm}(\Delta \hat{T}) &= a_{\rho}^\pm \left(1+b_{\rho}^{\pm} \Delta \hat{T}^{\beta}+c_{\rho}^{\pm} \Delta \hat{T}^{\Delta-1}+ \cdots \right)\\
&= A_\rho^{0,\pm} + A_\rho^{1,\pm} \Delta \hat{T}^{\beta} + A_\rho^{2,\pm}  \Delta \hat{T}^{\Delta-1} + \cdots\\
.
\end{split}
\label{}
\eeq
Based on the symmetric consideration, we should have $\theta^+$ =  $-\theta^-$. This gives,
\begin{equation}
\begin{split}
&\text{(i) asymmetric systems:} \\
&A_P^{0,+}/A_P^{0,-} = - 1,\quad A_P^{1,+}/A_P^{1,-}= 1,\quad A_P^{2,+}/A_P^{2,-} = -1,\\
&A_\rho^{0,+}/A_\rho^{0,-} = - 1,\quad A_\rho^{1,+}/A_\rho^{1,-} = 1,\quad A_\rho^{2,+}/A_\rho^{2,-} = 1 ,
\end{split}
\end{equation}
where we have used the properties that $p(\theta)$  and $s(\theta)$ are even functions~\cite{bertrand2011peculiar}. In general, for symmetric systems like the Ising model, naturally all terms obey the $Z_2$ symmetry,
\begin{equation}
\begin{split}
&\text{(ii) symmetric systems:} \\
&A_P^{0,+}/A_P^{0,-} = - 1,\quad A_P^{1,+}/A_P^{1,-}= -1,\quad A_P^{2,+}/A_P^{2,-} = -1,\\
&A_\rho^{0,+}/A_\rho^{0,-} = - 1,\quad A_\rho^{1,+}/A_\rho^{1,-} = -1,\quad A_\rho^{2,+}/A_\rho^{2,-} = -1.
\end{split}
\end{equation}
Thus, even though according to the Wegner expansion~\cite{wegner1976critical}, scaling corrections should generically appear in symmetric systems, the ``$+$'' and ``$-$'' terms  strictly cancel each other for the mean (diameter) of $L^\pm$ lines. In contrast, the correction terms can survive in asymmetric systems due to ``antisymmetry'' (such as $A_P^{1,+}/A_P^{1,-}=1$).
The above analysis reveals that the antisymmetric scaling corrections are the origin of asymmetric effects near the critical point. 

\vspace{10pt}  

\color{black}

\subsection{Linear mixing of two fields}

The asymmetry 
arises from two distinct sources: (i) the coupling between two ordering physical fields $P$ and $\mu$, (ii) the inclination ($\tan \varphi$) of the coexistence curve. To see that, let us consider two simpler cases with two-field mixing. 

First, if we assume that $b_2 = 0$ and $\tan\varphi = 0$, Eqs.~(\ref{eq:complete_scaling2}) and (\ref{eq:rho}) reduce to the following simplified forms: 
\beq
\begin{aligned}
h_{1} &= \Delta\hat{\mu} + a_{3} \Delta\hat{P}, \\
h_{2} &= \Delta\hat{T} , \\
h_{3} &= \Delta\hat{P} - \Delta\hat{\mu},
\label{eq:Pmucouple}
\end{aligned}
\eeq
and
\beq
\begin{aligned}
\Delta\hat{\rho} &\simeq (1 + a_3)\phi_1 + a_3(1 + a_3)\phi_1^2 .
\label{eq:rho_muP}
\end{aligned}
\eeq
This implies that the coupling between the two ordering fields introduces an asymmetric correction with a critical exponent of $2\beta$ since $\phi_1 \sim r^\beta$. 

On the other hand, if we simply assume that $a_1 = 0$ and $ b_2 = 0$, the forms of these two expressions reduce to: 
\beq
\begin{aligned}
h_{1} &=  a_{3} \tan\varphi \Delta\hat{T} + a_{3} \Delta\hat{P}, \\
h_{2} &= \Delta\hat{T} + \tan\varphi \Delta\hat{P}, \\
h_{3} &= \Delta\hat{P} - \Delta\hat{\mu} - \tan\varphi \Delta\hat{T},
\label{eq:PTcouple}
\end{aligned}
\eeq
and
\beq
\begin{aligned}
\Delta\hat{\rho} &\simeq  a_3\phi_1 + \tan\varphi\phi_2 .
\label{eq:rho_PT}
\end{aligned}
\eeq
This demonstrates that the tilt of the coexistence line introduces an asymmetric correction characterized by the critical exponent $1 - \alpha$ since $\phi_2 \sim r^{1-\alpha}$.

Taking the second scenario as an example, we select the critical exponents of the 3D Ising model, set $a_{3}=1$ and $a=k$ to simplify the parameters, and subsequently compute $L^{\pm}$. The result arising from the introduced asymmetry under these conditions is: 
\beq
\begin{split}
\frac{\Delta \hat{T}^-}{\Delta \hat{T}^+} &= 1 + k_P  |\delta \hat{P}|^{1 - \frac{1}{\Delta}},\\
\frac{\delta \hat{\rho}^-}{\delta \hat{\rho}^+} & = 1 + m_\rho |\delta \hat{P}|^{1 - \frac{1}{\Delta}}.
  \end{split}
    \label{eq:scaling_ratio_PT}
   \eeq
As illustrated in Fig.~\ref{fig:linear_theory}, the numerically obtained results agree well with the above expressions.

\begin{figure*}[!htbp]
  \centering
  \includegraphics[width=0.6\linewidth]{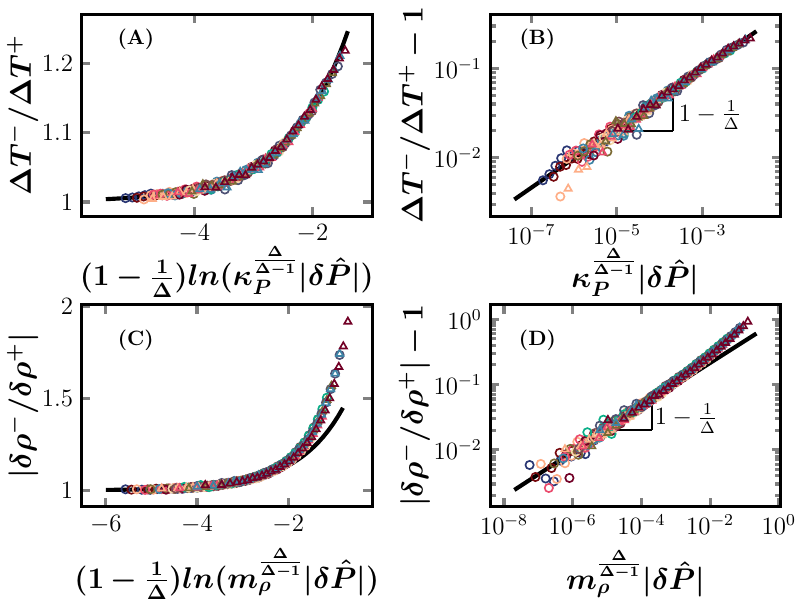}
  \caption{{\bf Asymmetric effects in $L^\pm$ lines examined by the linear scaling theory.}
  Data points are obtained from the EOSs given by the linear scaling theory, for $a=0.1$ ($\varphi = 30^{\circ}, 35^{\circ}, 40^{\circ}, 45^{\circ}, 50^{\circ}, 55^{\circ}, 60^{\circ}, 65^{\circ},  70^{\circ},  75^{\circ},  80^{\circ}$; circles), 
  and $\varphi = 82^{\circ}$ ($a = 0.02, 0.04, 0.06, 0.08, 0.10$; triangles). The critical exponents are set by the 3D Ising universality class values. The parameters $k_P$  and $m_\rho$, which depend on $a$ and $\varphi$, are obtained by fitting the data to Eq.~(\ref{eq:scaling_ratio_PT}). 
  The same data plotted in (A,B) semi-log and (C,D) log-log scales.
  The solid lines are $y = 1 + e^x$ and $y = x^{1-\frac{1}{\Delta}}$.
  }
  \label{fig:linear_theory}
\end{figure*}

\section{Higher-order cumulants of the particle number distribution}

\begin{figure}[!htbp]
  \centering
\includegraphics[width=0.9\linewidth]{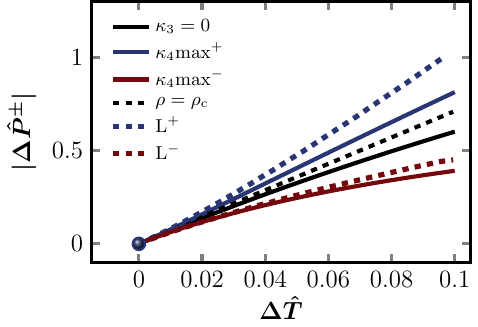}
  \caption{{\bf Comparison between the $\kappa_3 = 0$, maximum $\kappa_4$ lines and critical isochore, $L^\pm$ lines.}
  Data are obtained from the linear scaling theory with $a=0.1$ and $\varphi = 82^{\circ}$.
  }
  \label{fig:symmetry_line_compare}
\end{figure}

\begin{figure}[!htbp]
  \centering
  \includegraphics[width=0.75\linewidth]{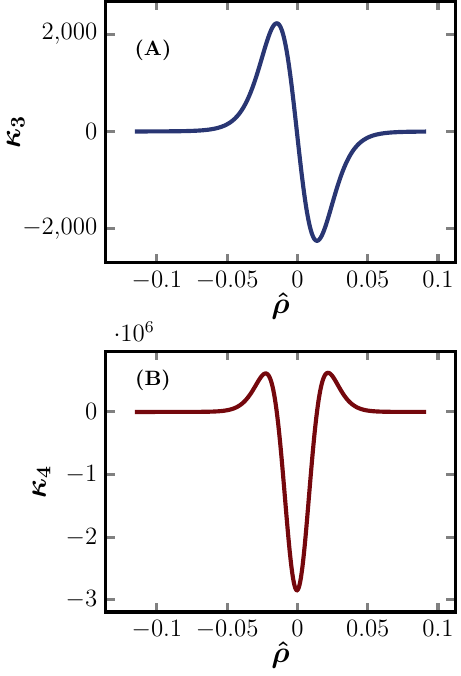}
  \caption{{\bf (A) $\kappa_3$ and (B) $\kappa_4$ as a function of $\hat{\rho}$ for a chosen $T$.} Data are obtained from the linear scaling theory with $a=0.1$, $\varphi = 30^{\circ}$ and $\Delta \hat{T} = 0.1$. 
  }
  \label{fig:rhoC111D1111}
\end{figure}

\begin{figure}[!htbp]
  \centering
  \includegraphics[width=0.8\linewidth]{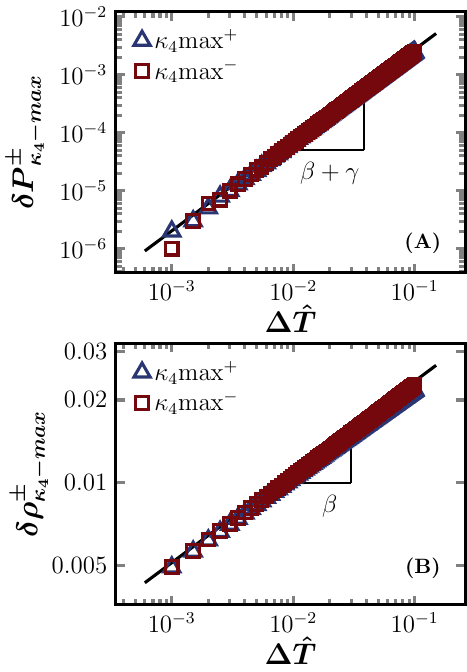}
  \caption{{\bf 
   The (A) pressure and (B) density differences  
  between $\kappa_3 = 0$ and $\kappa_4$-max lines.}  Data are obtained from the linear scaling theory with $a=0.1$ and $\varphi = 30^{\circ}$. 
  }
  \label{fig:kappa_lines}
\end{figure}

\begin{figure}[!ht]
  \centering
  \includegraphics[width=0.8\linewidth]{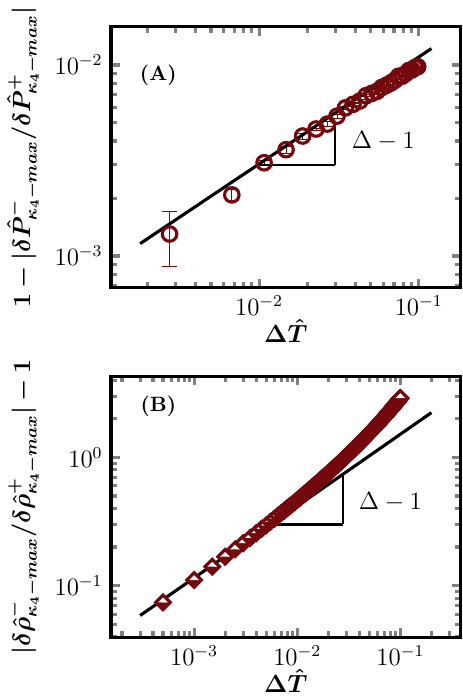}
  \caption{{\bf Asymmetric effects in the two  $\kappa_4$-max lines.} Data are obtained from the linear scaling theory with $a=0.1$ and $\varphi = 82^{\circ}$. 
  }
\label{fig:symmetry_line_asymmetry}
\end{figure}

We define intensive quantities corresponding to the cumulants $\kappa_n$ of the number distribution in a  grand canonical ensemble, $\kappa_2 = \mu_2/V$, $\kappa_3 = \mu_3/V$ and $\kappa_4 = (\mu_4 - 3 \mu_2^2)/V$, where $\mu_n = \langle (N- \langle N \rangle) ^{n} \rangle $ denotes the $n$-th central moment of particle number fluctuations. 
The fluctuation solution theory~\cite{ploetz2017fluctuation, ploetz2019gas} connects  $\kappa_n$  to macroscopic thermodynamic quantities with the following formulas,
\begin{equation}
\begin{aligned}
\kappa_{3} & = \frac{1}{\tilde{\beta}^{2}} \left[\rho_{}\left(\frac{\partial \rho_{}}{\partial P}\right)^{2}+\rho_{}^{2}\left(\frac{\partial^{2} \rho_{}}{\partial P^{2}}\right) \right],\\
\kappa_{4}
&=\frac{1}{\tilde{\beta}^{3}}\left[\rho_{}\left(\frac{\partial \rho_{}}{\partial \mathrm{P}}\right)^{3}+4 \rho_{}^{2}\left(\frac{\partial \rho_{}}{\partial \mathrm{P}}\right)\left(\frac{\partial^{2} \rho_{}}{\partial P^{2}}\right)+\rho_{}^{3} \frac{\partial^{3} \rho_{}}{\partial P^{3}}\right],
\end{aligned}
\end{equation}
where 
$\tilde{\beta}=(k_{B}T)^{-1}$ is the inverse temperature.

The EOSs are given by the linear scaling theory with the simplified two-field linear mixing of physical fields (see Eq.~(\ref{eq:scaling_ratio_PT})). In Fig.~\ref{fig:symmetry_line_compare}, the following lines are compared: the $L^\pm$ lines, the $\kappa_3 = 0$ line, and the two lines formed by the loci of $\kappa_4$ maxima. Note that   $\kappa_4(P)$ or $\kappa_4(\rho)$ has two maxima along an isothermal path~\cite{ploetz2019gas}, see Fig.~\ref{fig:rhoC111D1111}. 

It is interesting to examine if the characteristic lines defined by $\kappa_n$ exhibit the same scalings appearing in $L^{\pm}$. 
We have shown in Fig.~\ref{fig:rho_d} that the $\kappa_3 = 0$ line follows the asymmetric scaling Eq.~(\ref{eq:diameter_super}). Next we consider the $\kappa_3 = 0$ line as the reference line, and compute the pressure and density differences ($\delta P_{\kappa_4{\rm- max}}^{\pm}$ and $\delta \rho_{\kappa_4{\rm - max}}^{\pm}$) between the $\kappa_4$-max lines and the reference line, for a given $\Delta \hat{T}$. To the lowest order, these differences satisfy the scalings Eq.~(\ref{eq:scaling_LG}); see Fig.~\ref{fig:kappa_lines}. The next-order correction again shows the expected asymmetric scaling ($k_\rho = 0$ in Eq.~(\ref{eq:ratio}) due to two-field mixing; see Fig.~\ref{fig:symmetry_line_asymmetry}).
\vspace{180pt}

\section{Sensitivity to the critical temperature}

\begin{figure*}[!htbp]
\centering
\includegraphics[width=\linewidth]{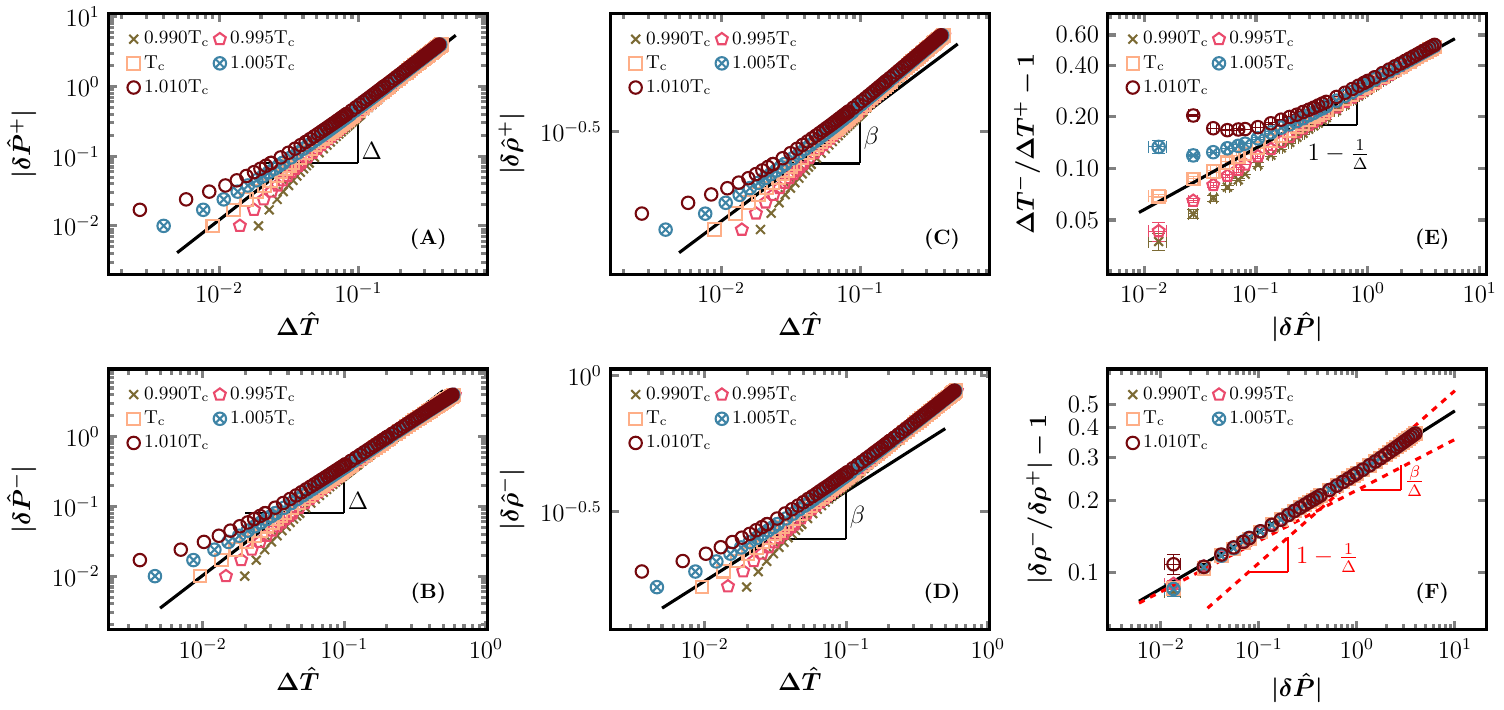}
\caption{{\bf Sensitivity analysis of the scaling relations to shifts in $T_{\rm c}$ for argon (Ar).} (A, B) $|\delta\hat{P}^{\pm}|$ as a function of $\Delta\hat{T}$. (C, D) $|\delta\hat{\rho}^{\pm}|$ as a function of $\Delta\hat{T}$. (E) $\Delta\hat{T}^{-}/\Delta\hat{T}^{+}-1$ as a function of $|\delta\hat{P}|$. (F) $|\delta\hat{\rho}^{-}/\delta\hat{\rho}^{+}|-1$ as a function of $|\delta\hat{P}|$. Different marks represent shifts of $T_{\rm c}$ by  $-1\%$ (circle), and $-0.5\%$ (circled times), $0\%$ (square), $0.5\%$ (pentagon), and $1\%$ (cross). The solid lines are the best fits using the unshifted  $T_{\rm c}$. The fitting  gives, in (A):~$y=16.0 x^{1.56}$; (B):~$y=13.5 x^{1.56}$; (C):~$y=0.75 x^{0.33}$; (D):~$y=0.80 x^{0.33}$; (E):~$y=0.30 x^{0.36}$; (F):~$y=0.19 x^{0.21} + 0.07 x^{0.36}$.}
\label{fig:Tc_sensitivity}
\end{figure*}

To examine how uncertainties in the critical temperature $T_{\rm c}$ affect the scaling analysis, we artificially shift $T_{\rm c}$ by $\pm0.5\%$ and $\pm1\%$ for argon (Ar), a representative fluid with well-characterized NIST data. For each shifted $T_{\rm c}$, we recompute the $L^{\pm}$ lines following the procedure in Appendix~A and then test the scaling relations in Eqs.~(5) and~(9). The results are summarized in Fig.~\ref{fig:Tc_sensitivity}. Panels (A) and (B) show the relation $\delta\hat{P}$ vs. $\Delta\hat{T}$, while panels (C) and (D) show $\delta\hat{\rho}$ vs. $\Delta\hat{T}$. Panels (E) and (F) display the two ratios in Eq.~(9): $\Delta\hat{T}^{-}/\Delta\hat{T}^{+}$ and $\delta\hat{\rho}^{-}/\delta\hat{\rho}^{+}$ as functions of $|\delta\hat{P}|$. 
All panels except (F) exhibit a dependence on $T_{\rm c}$.
The deviations observed in (E) directly originates from the deviations in the underlying scaling relations in (A, B). 

In general, one observes that the data with the original $T_{\rm c}$ (orange square) give the best agreement with the predicted scaling laws. Any shift of $T_{\rm c}$ introduces systematic deviations, especially in the near-critical region ($|\delta\hat{P}| \lesssim 0.1$). This indicates that our scaling analysis can, in principle, be used as a consistency check or even as a method to determine $T_{\rm c}$ from supercritical data: the value of $T_{\rm c}$ that minimizes deviations from the asymptotic scaling forms would be the optimal estimate. For $|\delta\hat{P}| \gtrsim 0.1$, the scaling relations remain robust against $\pm1\%$ shifts in $T_{\rm c}$, which is well above the typical uncertainty of $T_{\rm c}$ in the NIST database (generally lower than $0.1\%$ for the fluids studied). Therefore, our main conclusions are not compromised by $T_{\rm c}$ uncertainties, but we caution that the data points very close to $T_{\rm c}$ should be interpreted with care.

\end{document}